\documentclass[hidelinks,conference]{IEEEtran}

\usepackage{graphicx}
   \graphicspath{{./Figures/}}
\usepackage{footnote}   
\usepackage{amssymb}
\usepackage{amsmath}    
\usepackage{algorithm}
\usepackage{algorithmic}
\usepackage{multirow}
\usepackage{svg}
\usepackage{color}
\DeclareGraphicsExtensions{.pdf,.jpeg,.png,.fig, .emf}
   \usepackage{subfigure}
    \usepackage{epstopdf}
    \usepackage{epsfig}
    \usepackage{subfigure}
    \usepackage{epstopdf}
    \usepackage{epsfig}

\usepackage{RobStd}
\HideNotes
\usepackage{makecell}
\usepackage{booktabs}
\usepackage{multirow}
\usepackage{xspace}
\newcommand{\design}{SEE-MCAM\xspace}
\usepackage{makecell}
\usepackage{siunitx}
\newcommand{\shengxi}[1]{\textcolor{blue}{#1}}
\newcommand{\chekai}[1]{\textcolor{magenta}{#1}}
\newcommand\eat[1]{}
\newcommand{\overbar}[1]{\mkern 1.5mu\overline{\mkern-1.5mu#1\mkern-1.5mu}\mkern 1.5mu}

\begin{document}
\bstctlcite{IEEEexample:BSTcontrol}
\title{SEE-MCAM: Scalable Multi-bit FeFET Content Addressable Memories for Energy Efficient Associative Search}
 \author{
 \small
 Shengxi Shou$^{1,3}$, Che-Kai Liu$^{2}$, Sanggeon Yun$^3$, Zishen Wan$^{2}$, Kai Ni$^4$, Mohsen Imani$^3$, X. Sharon Hu$^5$,\\ Jianyi Yang$^6$, Cheng Zhuo$^{6*}$, Xunzhao Yin$^{1,7*}$\\
 $^1$College of Information Science and Electronic Engineering, Zhejiang University, China\\
$^2$School of Electrical and Computer Engineering, Georgia Institute of Technology, USA\\
 $^3$Department of Information and Computer Science, University of California Irvine, USA\\
 $^4$Department of Electrical and Microelectronic Engineering, Rochester Institute of Technology, USA\\
 $^5$Department of Computer Science and Engineering, University of Notre Dame, USA\\
 $^6$School of Micro-Nano Electronics, Zhejiang University, China\\
 $^7$Key Laboratory of Collaborative Sensing and
Autonomous Unmanned Systems of Zhejiang Province, China\\
$^*$Corresponding authors, email: \{czhuo, xzyin1\}@zju.edu.cn
\vspace{-1ex}
}

\maketitle
\pagestyle{empty}

%

\begin{abstract}

Artificial intelligence has made remarkable advancements in recent years, leading to the development of algorithms and models capable of handling ever-increasing amounts of data. The computational demands of these algorithms necessitate circuit and architecture designs that go beyond the von-Neumann paradigm.
Content addressable memories (CAMs), which implement parallel associative search functionality within memory blocks to overcome the memory wall bottleneck, have proven to be effective for data-intensive tasks. While current CAM designs have achieved higher storage density and energy efficiency than their CMOS-based counterparts by leveraging emerging non-volatile memories (NVM), most of these implementations are limited to binary storage cells.
In this work, we propose SEE-MCAM, scalable and compact \textit{multi-bit CAM (MCAM)} designs that utilize the three-terminal ferroelectric FET (FeFET) as the proxy. 
By exploiting the multi-level-cell characteristics of FeFETs, 
our proposed SEE-MCAM designs enable multi-bit associative search functions and achieve better energy efficiency and performance than existing FeFET-based CAM designs. 
We validated the functionality of our proposed designs by achieving 3 bits per cell CAM functionality, resulting in 3$\times$ improvement in storage density. The area per bit of the proposed SEE-MCAM cell is 8\%  of the conventional CMOS CAM.
We thoroughly investigated the scalability and robustness of the proposed design. 
Evaluation results suggest that the proposed 2FeFET-1T \design achieves 9.8$\times$ more energy efficiency and 1.6$\times$ less search latency compared to the CMOS CAM, respectively. 
When compared to existing MCAM designs, the proposed \design can achieve 8.7$\times$ and 4.9$\times$ more energy efficiency than  ReRAM-based and FeFET-based MCAMs, respectively.
Benchmarking results show that our approach provides up to 3 orders of magnitude improvement in speedup and energy efficiency over a GPU implementation in accelerating a novel quantized hyperdimensional computing (HDC) application.
\end{abstract}

\section{Introduction}
\label{sec:introduction}

In the era of artificial intelligence (AI), the exponential growth of data generated by machine learning applications, edge devices, and data centers has created significant demands on the efficiency of the underlying hardware. Such hardware needs to support high-performance and data-intensive applications. One crucial operation in these algorithms is the data query operation, which involves searching for a vector among a large number of data vectors stored in a library. This operation is integral to various machine learning and neuromorphic models, such as hyperdimensional computing (HDC) \cite{imani2017exploring,zou2022biohd}, few-shot learning \cite{hersche2022constrained}, reinforcement learning \cite{li2022associative}, bioinformatics \cite{barkam2023hdgim}, and robotics \cite{wan2021survey,  shi2020acoustic}.


However, traditional Von-Neumann hardware faces a challenge known as the memory wall problem when handling these data-intensive workloads. The memory wall problem arises due to the substantial data movement required between the memory and computing units, resulting in significant data transfer overhead. This overhead dominates the total cost of data query operations, ultimately limiting the overall efficiency of the system.

To address this challenge and enhance efficiency, there is a strong demand for hardware solutions that support parallel associative search (i.e., data query operations) within the memory, effectively eliminating data transfer overhead. In-memory computing (IMC) has emerged as an alternative architectural paradigm that combines computational and storage units, offering promising solutions to overcome the memory wall challenge specifically for data search operations. Content addressable memory (CAM) is a key primitive of IMC, embedding parallel search functionality within memory blocks and enabling fast associative search. CAM has gained significant adoption as an associative memory (AM) to accelerate the inference phase of novel machine learning tasks mentioned earlier.

Conventional CMOS-based 16T CAM arrays \cite{pagiamtzis2006content} suffer from drawbacks such as high leakage and area overhead due to the energy-consuming 6T static random access memory (SRAM) design. Recent research focuses on leveraging emerging non-volatile memories (NVMs) such as ferroelectric field-effect transistors (FeFETs) \cite{ni2018circuit, ni2019ferroelectric}, resistive random access memory (ReRAM) \cite{wong2012metal, chang20173t1r, li20131}, and spin-transfer torque magnetic RAM (STT-MRAM) \cite{garzon20234} to develop more compact and efficient CAM designs. By storing binary logic values inside NVM devices and performing bit-wise XOR logic operations between the query and the stored data, NVM-based binary CAM (BCAM) and ternary CAM (TCAM) designs have proven to be more compact and energy-efficient than their CMOS counterparts. These NVM-based CAM designs have been extensively studied under various data-intensive workloads \cite{imani2017exploring, hersche2022constrained, li2022associative, barkam2023hdgim}.

However, these NVM-based CAM designs have been limited to exploiting the single-level cell (SLC) property of NVM, hindering further improvements in CAM density. To overcome this limitation and  enhance CAM density, leveraging the \textit{multi-level cell (MLC)} property of NVMs for \textit{multi-bit CAMs (MCAMs)} has become an appealing direction of research. Some approaches include  a 6T-2R MCAM  utilizing the MLC property of ReRAM devices, but it requires  additional transistors and exhibits energy consumption due to  analog inverters and  current-based sensing \cite{li2020analog, pedretti2021tree}. FeCAM  in \cite{yin2020fecam}, achieves MCAM functionality using only two FeFET devices, but introduces high precharge energy associated with the CAM matchline (ML). A 2FeFET-1T CAM design \cite{li2020scalable} eliminates  analog inverters but requires separate sensing circuitry for NOR-type and NAND-type ML structures,  and  is vulnerable to device variations. Additionally, a  distance function based on  FeFET conductance within the MCAM array \cite{kazemi2021fefet}  faces challenges related to FeFET variations.

To fully exploit the potential of NVM-based CAMs for accelerating data-intensive workloads in-memory, it is crucial to design CAMs that effectively address the aforementioned drawbacks of existing works. Such a CAM design could lead to significantly higher CAM density and improved performance and energy efficiency, while incurring minimal penalties in terms of robustness.

In this work, we propose SEE-MCAM, which leverages FeFETs as proxy NVMs, to design scalable NOR-type and NAND-type MCAMs for energy-efficient associative search. Using an experimentally calibrated FeFET Presaich model \cite{ni2018circuit}, we incorporate a 2FeFET multi-bit input binary output (MIBO) XOR logic  structure into our proposed CAM cells, allowing for storage of multi-bit values and implementation of the Boolean XOR logic operation. Such structure effectively controls the access transistors associated with the CAM array matchline (ML), reducing or eliminating the precharge energy typically associated with the ML and ensuring robustness against device variations. Additionally, our proposed CAM array can be programmed by employing write inhibition schemes and directly applying multi-bit search operations to the FeFET source/drain \cite{ni2018write, xiao2022write}. Building upon the 2FeFET MIBO XOR logic structure, we propose NOR-type 2FeFET-1T and NAND-type 2FeFET-2T SEE-MCAM designs that support multi-bit and parallel search functions. 
The NOR-type SEE-MCAM reduces ML-associated capacitance, resulting in energy savings, while the NAND-type SEE-MCAM eliminates the precharge phase.

We discuss and evaluate the structures, operations, simulation validation, and energy/performance analysis of the proposed SEE-MCAM designs at the array level. 
To demonstrate the advantages of utilizing FeFETs' MLC property in conjunction with SEE-MCAM design schemes, we compare our designs with existing MCAM designs based on ReRAM and FeFET, respectively. Additionally, we investigate the scalability and robustness of SEE-MCAM arrays against device variations. 
We have demonstrated that SEE-MCAM  achieves up to 3 bits per cell CAM density. 
With a simpler cell structure and sensing circuitry, the SEE-MCAM array significantly improves energy efficiency compared to prior works \cite{li2020analog, li2020scalable, qian2021energy, yin2022ferroelectric, ni2019ferroelectric}. Specifically, SEE-MCAM demonstrates an area per bit efficiency of 8\%  of CMOS CAM and achieves 8.7$\times$ and 4.9$\times$ higher energy efficiency than ReRAM and FeFET-based MCAMs, respectively, while maintaining sufficient robustness against device variations.
Furthermore, benchmarking results of a novel quantized HDC inference task using the SEE-MCAM array indicate a potential improvement of up to 3 orders of magnitude compared to conventional GPU-based approaches.

\section{Background}
\label{sec:background}
\subsection{FeFET Basics}
\label{sec:device}
\begin{figure}
    \centering
    \includegraphics[width=\linewidth]{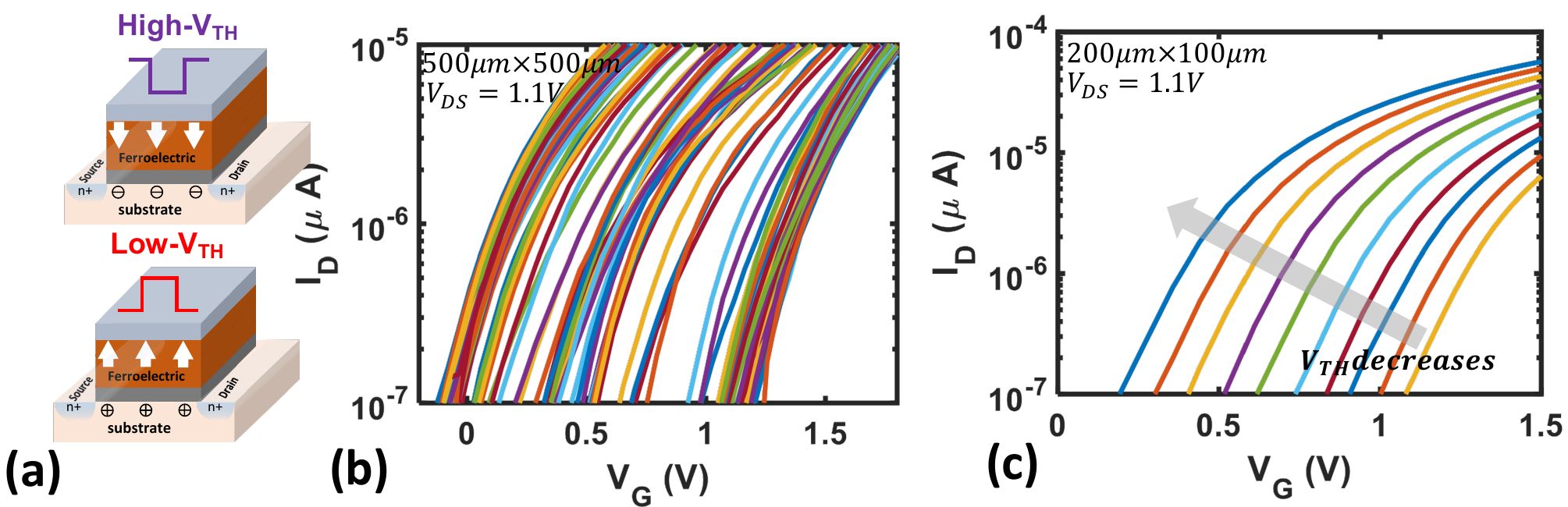}
    \caption{(a)  FeFET  write schemes. (b)  $I_{D}-V_{G}$ characteristics with MLC $V_{TH}$ states of a fabricated FeFET. (c) Simulated $I_{D}-V_{G}$ characteristics with more than 3-bit $V_{TH}$ statesbased on Preisach FeFET model. }
    \label{fig:fefet}
\end{figure}

FeFETs based on HfO$_2$ \cite{khan2020future} have emerged as highly competitive candidates due to their intrinsic CMOS structure, high $I_{ON}/I_{OFF}$ ratio, low OFF current, excellent scalability, CMOS compatibility, and superior write energy efficiency. FeFETs are fabricated by integrating a ferroelectric layer in the gate stack of a metal-oxide-semiconductor field-effect transistor (MOSFET), where HfO$_2$ serves as the ferroelectric material (as shown in Figure \ref{fig:fefet}(a)). Moreover, Figure \ref{fig:fefet}(a) illustrates a FeFET that can store high-$V_{TH}$ and low-$V_{TH}$ states, respectively. By applying a positive (negative) voltage pulse to the gate terminal, the polarization of the ferroelectric layer will be switched towards the channel, programming the FeFET into the low-$V_{TH}$ (high-$V_{TH}$) state. Figure \ref{fig:fefet}(b) and (c) show the $I_{D}-V_{G}$ characteristics of the fabricated and simulated FeFET devices with different write pulses, respectively.
FeFETs have been successfully deployed in various scenarios, including CAMs \cite{yin2022ferroelectric}, frequency multipliers \cite{mulaosmanovic2020reconfigurable}, crossbars \cite{liu2022cosime}, field-programmable gate arrays (FPGAs) \cite{chen2018power}, and oscillators \cite{fang2019neuro}, among others.

\subsection{Existing Binary CAMs and Ternary CAMs}

\begin{figure}
    \centering
    \includegraphics[width=\linewidth]{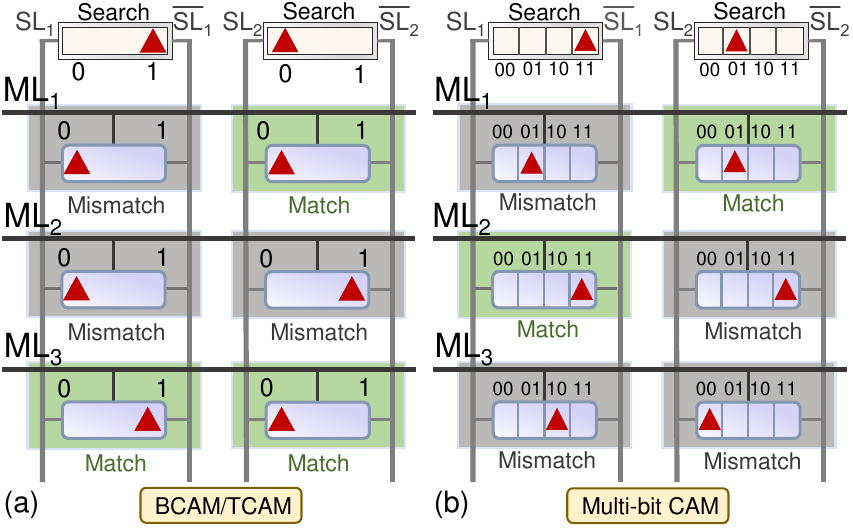}
    \caption{(a) In a single bit CAM, a '0' or '1' is stored and searched in parallel, while (b) in a MCAM, multiple level values can be stored and searched in parallel.}
    \label{fig:CAM_compare}
\end{figure}

\begin{figure}
    \centering
    \includegraphics[width=\linewidth]{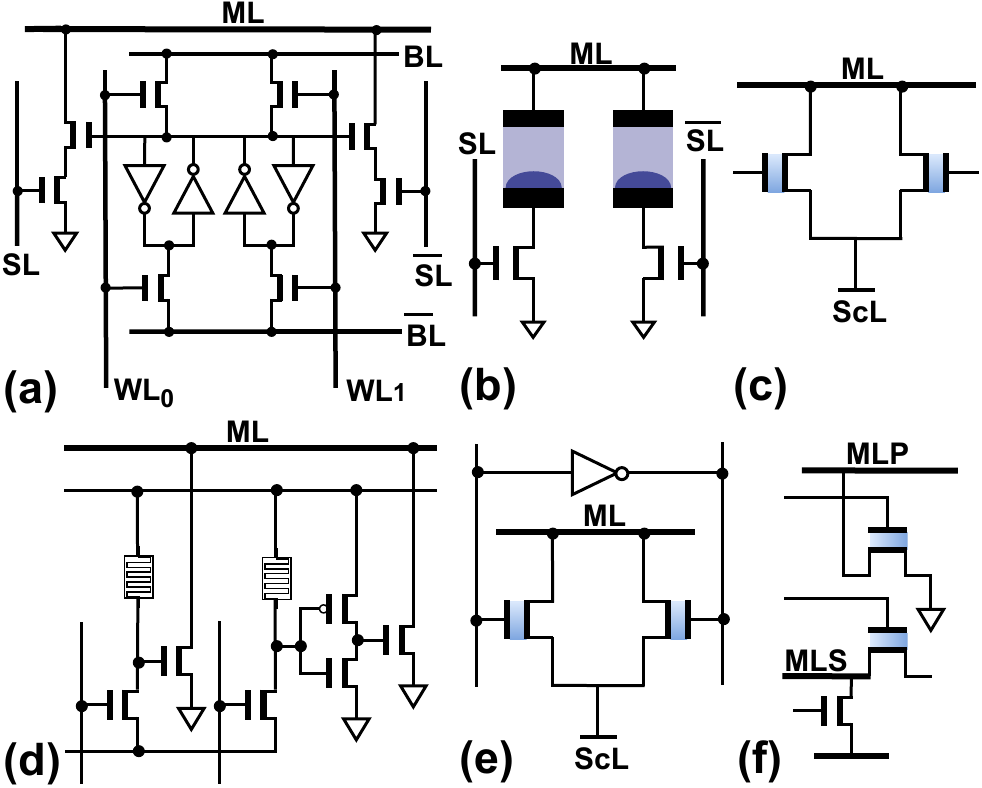}
    \caption{Existing BCAM/TCAM designs based on (a) 16T CMOS \cite{pagiamtzis2006content}, (b) 2T-2R \cite{li20131}, and (c) 2FeFET \cite{ni2019ferroelectric}. Existing MCAM designs based on (d) 6T-2ReRAM \cite{li2020analog}, (e) 2FeFET \cite{yin2020fecam}, and (f) 2FeFET-1T \cite{li2020scalable}.}
    \label{fig:existing_CAM}
\end{figure}

Various CAM designs have been proposed based on canonical CMOS and different NVM devices. 
Depending on the type of data stored in a CAM cell, CAMs can be categorized into binary CAM (BCAM), ternary CAM (TCAM), multi-bit CAM (MCAM), and analog CAM (ACAM) \cite{hu2021memory}. Most of the existing CAM designs are BCAM, with binary values stored in cell, implementing bit-wise XOR logic. TCAM can store an additional "don't care" bit besides binary values, which serves as a wildcard. \autoref{fig:CAM_compare} illustrates the  BCAM and TCAM storing a single bit.
\autoref{fig:existing_CAM} summarizes some representative BCAM and TCAM designs. \autoref{fig:existing_CAM}(a) presents the traditional 16T CMOS-based CAM, which consumes significant energy and area overhead.  
ReRAM-based CAMs such as the 2T-2R TCAM \cite{li20131} (\autoref{fig:existing_CAM}(b)) and 3T-1R CAM \cite{chang20173t1r} have been proposed and fabricated for memory-intensive tasks \cite{li2016looking}.  
Though these design consume much less area overhead than CMOS, the low HRS/LRS (high/low resistance state) ratio, current driven write mechanisms and  two-terminal NVM structure of ReRAM devices necessitate extra selectors and write facilitation circuitry, thus resulting in high energy consumption \cite{ni2019ferroelectric}. 
Recently, FeFET emerges as a promising device due to its high $I_{ON}/I_{OFF}$ ratio, low $I_{OFF}$ and three-terminal structure.     A number of FeFET-based CAMs have been  proposed for energy-efficient  data-intensive computing tasks \cite{tan2019experimental, ni2019ferroelectric, yin2022ferroelectric, cai2022energy, liu2023reconfigurable}.
\autoref{fig:existing_CAM}(c) depicts a typical 2FeFET CAM cell \cite{yin2018ultra}. That said, these FeFET designs only exploit the binary NVM property of FeFETs. As shown in \autoref{fig:fefet}, the potential of FeFETs remains unexplored.

\eat{ Multi-level cell (MLC) CAM has recently gained its attention \cite{li2020scalable}, but at the circuit level, the peripheries may still incur a large overhead with two CAM branches representing a single vector. In addition, current multi-bit best-match CAM designs are reported at most 2-bit \cite{li2020scalable}. To overcome the above challenges, based on the recently proposed 2FeFET-1T CAM \cite{yin2022ferroelectric}, we proposed an ultra-efficent CAM that is capable of storing up to 3-bit per cell.}

\subsection{MCAM Concepts and Related Works}
\label{sec:MCAM_works}

Above NVM-based BCAMs and TCAMs are limited to exploiting the SLC characteristic of NVMs, thus hindering from further CAM density improvement. 
Recent works explore the possibilities of exploiting the MLC properties of NVMs to construct MCAM designs to boost the CAM density.
Unlike the single-bit CAM design shown in \autoref{fig:CAM_compare}(a), MCAMs store multi-bit values, and a multi-bit input query is applied for a search. Only when all multi-bit values in the query are identical to a stored entry, a match can be detected.
\cite{li2020analog} and  \cite{pedretti2021tree} presented a 6T-2R MCAM (\autoref{fig:existing_CAM}(d)) that utilizes the MLC property of ReRAM devices, albeit at the cost of four additional transistors compared to the conventional 2T-2R TCAM design \cite{li20131}. Moreover, their design incorporates an analog inverter and a current-based sensing mechanism, resulting in significant energy consumption. \cite{yin2020fecam} proposed FeCAM, which employs only two FeFET devices to achieve MCAM functionality, shown in \autoref{fig:existing_CAM}(d). However, FeCAM associates the two FeFETs' drain capacitance with the CAM matchline (ML), introducing high precharge energy. Another approach, the 2FeFET-1T CAM introduced in  \cite{li2020scalable}, eliminates the need for analog inverters by dividing the two FeFETs into separate NOR-type and NAND-type ML branches, shown in \autoref{fig:existing_CAM}(f). Unfortunately, the two ML branches require different sensing circuitry, and the NAND-type branch is vulnerable to device variations.
\cite{kazemi2021fefet} and \cite{kazemi2022achieving} propose to realize a novel distance function within an MCAM cell and the cell structure is the same as that of 2FeFET TCAM in \autoref{fig:existing_CAM}(c). However, such a distance function relies on accurate FeFET conductance in the linear and saturation region, which makes it vulnerable to FeFET variations.
Our proposed SEE-MCAM designs aim to exploit the potential of FeFET devices, achieving higher CAM density, improved performance and energy efficiency than the above works, while maintaining the robustness.

\section{Proposed SEE-MCAM Designs}
\label{sec:schematic}
In this section, we propose two SEE-MCAM designs with multi-bit functionality and improved energy efficiency by either (i) reducing the NOR-type ML capacitance, or (ii) eliminating the precharge phase in NAND-type ML. We first introduce the 2FeFET structure implementing the key MIBO XOR logic, and then discuss the SEE-MCAM designs.

\subsection{2FeFET Structure for Multi-Bit Input Binary Output}
\label{sec:MIBO}

\begin{figure}
    \centering
    \includegraphics[width=\linewidth]{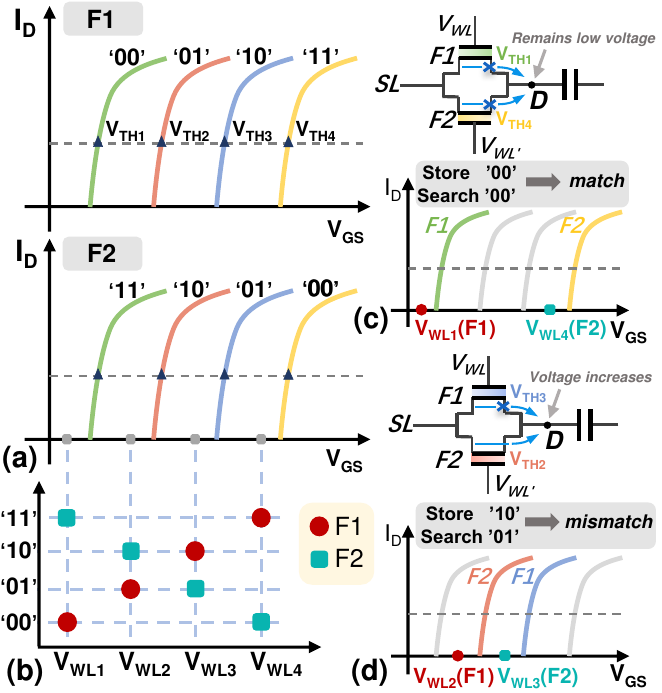}
    \caption{A 2bit case of 2FeFET MIBO: (a) F1 and F2 are written to store 4 levels of threshold voltage states, respectively. (b) 4 sets of $V_{\text{WL}}$  applied at the gates of F1 and F2 correspond to search 4 different values, respectively.   (c) A match case where  the stored value is `00', and the input value is `00'. The binary output $D$ is at low. (d) A mismatch case where the stored value `10',  and the input value is `01'. The binary output $D$ is at high. }
    \label{fig:Schematic}
\end{figure}

\autoref{fig:Schematic} shows the schematic and an example 2-bit operation principle of the 2FeFET structure to implement MIBO XOR logic, which is the key function in a CAM cell.
The 2FeFETs are connected in parallel, forming a push-pull structure. By programming multiple threshold voltages, i.e., $V_{TH1}$, $V_{TH2}$, $V_{TH3}$, $V_{TH4}$, into the 2FeFETs  F1 and F2 to encode the stored multi-bit value as shown in \autoref{fig:CAM_compare}(a), and then applying pre-defined search voltages, i.e., $V_{WL1}$, $V_{WL2}$, $V_{WL3}$, $V_{WL4}$, corresponding to the search query values shown in \autoref{fig:Schematic}(b), a binary XOR output (match or mismatch) between the stored value and the applied query input can be generated. During the operation, The sourceline ($SL$) is pulled up to high level. 


\autoref{fig:Schematic}(c) and (d) demonstrate a match and a mismatch cases, respectively.
Consider a 2FeFET cell storing `00' as shown in \autoref{fig:Schematic}(c), $F1$ stores $V_{TH1}$ and $F2$  stores $V_{TH4}$. 
Then an input value `00' is searched by applying corresponding $V_{WL}$ set per \autoref{fig:Schematic}(b), i.e., applying $V_{WL1}+V_{SL}$ to the gate of $F1$, and $V_{WL4}+V_{SL}$ to the gate of $F2$, respectively. 
In this way, since both FeFETs are turned off, the output node $D$ is not charged by $SL$, and remains at low level, indicating a match case.
Similarly, \autoref{fig:Schematic}(d) demonstrates a 2FeFET cell storing "10", where $F1$ stores $V_{TH3}$ and $F2$ stores $V_{TH2}$.An input query "01" is searched by applying $V_{WL2}+V_{SL}$ to $F1$ and $V_{WL3}+V_{SL}$ to $F2$, respectively. In this case, $F1$ is turned off but $F2$ turns on. The output node $D$ will then be charged to high level, indicating a mismatch case. It can be seen that the 2FeFET structure implements a MIBO XOR logic, only when the input query is identical to the stored value, the output is at low and high otherwise. This MIBO principle is scalable depending on the number of distinguishable states a FeFET can store, and 3-bit per cell CAM function will be discussed in this section.
This 2FeFET structure forms the basic of our proposed SEE-MCAM designs.

\subsection{2FeFET-1T SEE-MCAM}
\label{sec:2FeFET-1T}

\begin{figure}
    \centering
    \includegraphics[width=\linewidth]{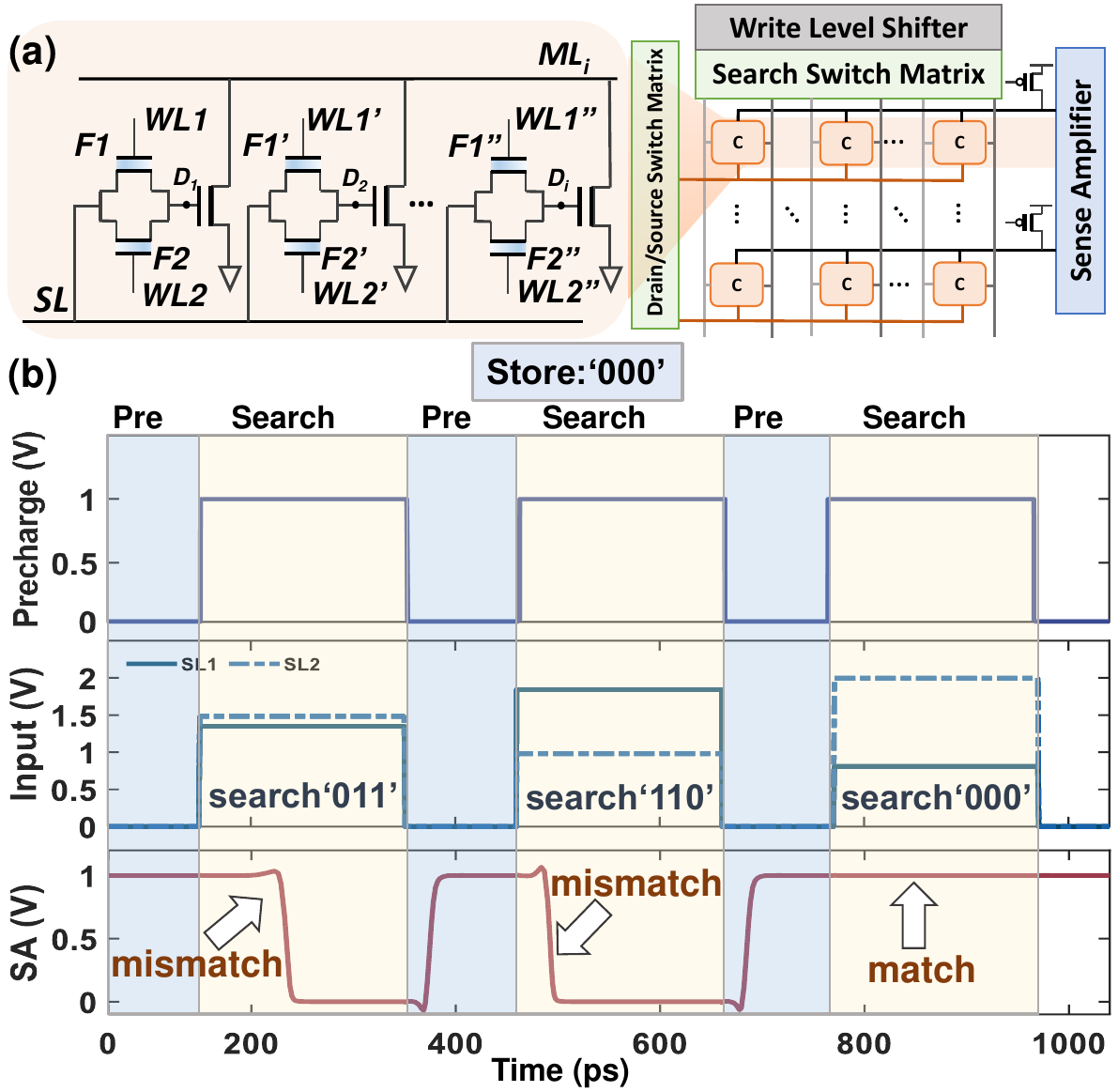}
    \caption{The proposed NOR type SEE-MCAM design. (a) Schematic of the 2FeFET-1T SEE-MCAM array. (b) Transient waveform of a 3-bit 2FeFET-1T SEE-MCAM.}
    \label{fig:wave}
\end{figure}

By embedding the 2FeFET structure in Sec. \ref{sec:MIBO} into the CAM cell as shown in 
\autoref{fig:wave}(a), the proposed NOR type 2FeFET-1T SEE-MCAM is realized.  
\autoref{fig:wave}(a)
shows the schematic of 2FeFET-1T SEE-MCAM design. A NMOS access transistor separates the 2FeFET structure from ML.  The wordlines $WL1$ and $WL2$ shared by each column control the gates of 2FeFET structure, and the $SL$ shared by a row connects the sources of the 2FeFET structures within the word. The MIBO node $D$ is connected to the gate of NMOS. The sense amplifier (SA) of the array adopts a threshold inverter quantization (TIQ) comparator \cite{cai2022energy}.

During the write, the $SL$s associated with unselected words are applied with write inhibition scheme \cite{ni2018write, xiao2022write}.
The $SL$s of selected words as well as the wordlines of selected columns are applied with write pulses to program the FeFETs into desired threshold voltage state per \autoref{fig:fefet}.
\autoref{tab:operation} summarizes the search operations of the 2FeFET-1T SEE-MCAM cell. The MLs of the array are precharged to a high level. Due to the NOR type ML connection, only when all input values are identical to the stored values of the cells within a word, the corresponding word ML can remain at high, otherwise drops down to a low level. 
\autoref{fig:wave}(b) shows the transient waveforms that validate the 3-bit MCAM function of NOR type \design.

\begin{table}[]
\centering
\caption{Operation summary of a 3-bit SEE-MCAM.}
\label{tab:operation}
\resizebox{1\columnwidth}{!}{
\begin{tabular}{|c|cccccccc|}
\toprule
\shortstack{\textbf{$D$}\\\textbf{ML}}& \shortstack{\textbf{Stored}\\ \textbf{`000'}} & \shortstack{\textbf{Stored}\\ \textbf{`001'}}  & \shortstack{\textbf{Stored}\\ \textbf{`010'}}  & \shortstack{\textbf{Stored}\\ \textbf{`011'}} & \shortstack{\textbf{Stored}\\ \textbf{`100'}}  & \shortstack{\textbf{Stored}\\ \textbf{`101'}}  & \shortstack{\textbf{Stored}\\ \textbf{`110'}}  & \shortstack{\textbf{Stored}\\ \textbf{`111'}}  \\
\midrule
\shortstack{\textbf{Search}\\\textbf{`000'}}&  \makecell{\textit{L}\\\textit{M}} &  \makecell{H\\MM} &  \makecell{H\\MM} & \makecell{H\\MM}&\makecell{H\\MM}&\makecell{H\\MM}&\makecell{H\\MM}&\makecell{H\\MM} \\ \midrule
\shortstack{\textbf{Search}\\\textbf{`001'}}&  \makecell{H\\MM} &  \makecell{\textit{L}\\\textit{M}} &  \makecell{ H\\ MM} & \makecell{H\\MM}&\makecell{H\\MM}&\makecell{H\\MM}&\makecell{H\\MM}&\makecell{H\\MM} \\ \midrule
\shortstack{\textbf{Search}\\\textbf{`010'}}&  \makecell{H\\MM} &  \makecell{H\\MM} &  \makecell{ \textit{L}\\ \textit{M}} & \makecell{ H\\ MM}&\makecell{H\\MM}&\makecell{H\\MM}&\makecell{H\\MM}&\makecell{H\\MM} \\ \midrule
\shortstack{\textbf{Search}\\\textbf{`011'}}&  \makecell{H\\MM} &  \makecell{H\\MM} &  \makecell{ H\\ MM} & \makecell{ \textit{L}\\ \textit{M}}&\makecell{H\\MM}&\makecell{H\\MM}&\makecell{H\\MM}&\makecell{H\\MM} \\ \midrule
\shortstack{\textbf{Search}\\\textbf{`100'}}&  \makecell{H\\MM} &  \makecell{H\\MM} &  \makecell{H\\MM} & \makecell{ H\\ MM}&\makecell{\textit{L}\\\textit{M}}&\makecell{H\\MM}&\makecell{H\\MM}&\makecell{H\\MM} \\ \midrule
\shortstack{\textbf{Search}\\\textbf{`101'}}&  \makecell{H\\MM} &  \makecell{H\\MM} &  \makecell{H\\MM} & \makecell{H\\MM}&\makecell{H\\MM}&\makecell{\textit{L}\\\textit{M}}&\makecell{H\\MM}&\makecell{H\\MM} \\ \midrule
\shortstack{\textbf{Search}\\\textbf{`110'}}&  \makecell{H\\MM} &  \makecell{H\\MM} &  \makecell{H\\MM} & \makecell{H\\MM}&\makecell{H\\MM}&\makecell{H\\MM}&\makecell{\textit{L}\\\textit{M}}&\makecell{H\\MM} \\ \midrule
\shortstack{\textbf{Search}\\\textbf{`111'}}&  \makecell{H\\MM} &  \makecell{H\\MM} &  \makecell{H\\MM} & \makecell{H\\MM}&\makecell{H\\MM}&\makecell{H\\MM}&\makecell{H\\MM}&\makecell{\textit{L}\\\textit{M}} \\ \midrule

\end{tabular}
}
\flushleft
For a given stored and input value pair. H/L indicates the high/low voltage level of the node $D$. M/MM indicates a match or mismatch state of ML.
\end{table}

Compared with prior FeFET-based CAM designs \cite{qian2021energy, yin2022ferroelectric, li2020scalable, yin2020fecam}, our proposed 2FeFET-1T SEE-MCAM addresses their respective drawbacks,  thus achieving the multi-bit search operation with significant energy efficiency.
The BCAM/TCAM design from \cite{qian2021energy, yin2022ferroelectric} is limited to single-bit CAM functionality, and requires two complementary $SL$s associated with the CAM cell to perform the search. Our proposed SEE-MCAM, on the other hand, increases the data density without any additional transistors and peripherals, and only requires one supply rail $SL$ to perform the search operation.
The MCAM designs from \cite{ yin2020fecam} consumes significant precharge energy as it associates two FeFETs with ML, resulting in a large ML capacitance to be precharged:
\begin{equation}
\label{eq:normal_delay}
\begin{aligned}
    C_{\text{ML}}& \approx C_{\text{d,P}} + N\times(2C_{\text{FeFET}}+C_{\text{parasitic}})\\
\end{aligned}
\end{equation}
where $C_{\text{d,P}}$, $C_{\text{parasitic}}$ and $C_{\text{FeFET}}$ are the drain capacitance of the precharge PMOS, the parasitic capacitance associated with ML, and the drain capacitance of the FeFET, respectively. $N$ is the number of cells within a word. 
The 2FeFET-1T MCAM from \cite{li2020scalable} consumes the same device count as the proposed design, but the two ML branches of the MCAM design incur both high precharge energy and  latency. 
Our proposed 2FeFET-1T SEE-MCAM array reduces the transistor associated with ML to only 1, therefore consumes less precharge energy while maintaining the latency:
\begin{equation}
\label{eq:2fefet-1t}
\begin{aligned}
    C_{\text{ML}}& \approx C_{\text{d,P}} + N\times(C_{\text{NMOS}}+C_{\text{parasitic}})\\
\end{aligned}
\end{equation}

\subsection{Precharge-Free 2FeFET-2T SEE-MCAM}
\label{sec:prefree}

\begin{figure}
    \centering
    \includegraphics[width=\linewidth]{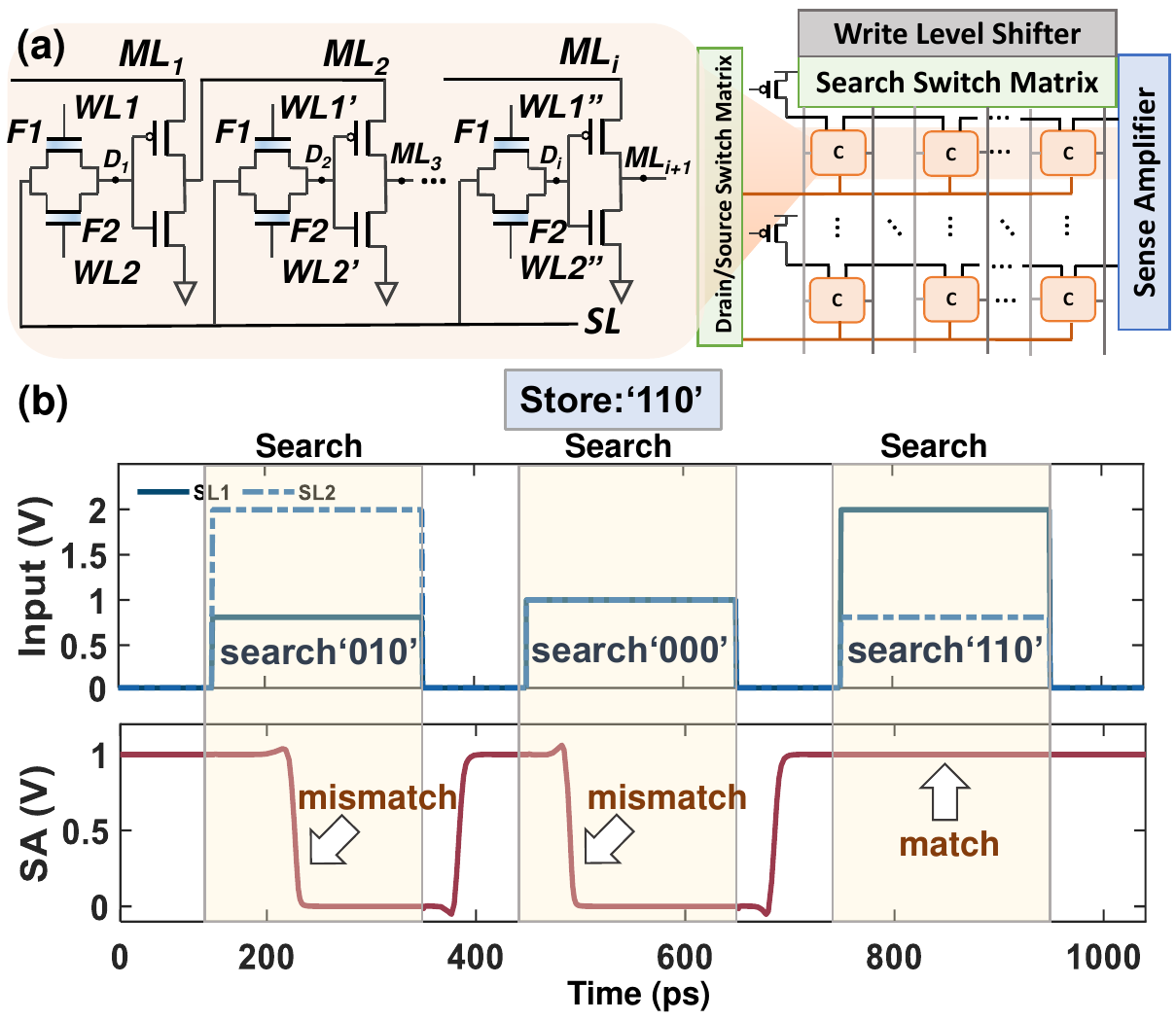}
    \caption{Precharge-free multibit CAM. (a) Schematic of 2FeFET-2T precharge-free MCAM. (b) Transient waveforms of the 3-bit 2FeFET-2T precharge-free MCAM. }
    \label{fig:prefree}
\end{figure}

To further improve the energy efficiency of SEE-MCAM, we propose a NAND type 2FeFET-2T SEE-MCAM array that embeds the 2FeFET MIBO structure and leverages a  precharge-free scheme \cite{yin2022ferroelectric} to eliminate the energy consuming precharge phase.
\autoref{fig:prefree}(a) shows the schematic of the proposed 2FeFET-2T SEE-MCAM cell and array, respectively. The cell consists of the 2FeFET MIBO structure and an inverter. The array ML  adopts the NAND type connection, where the ML of the previous cell is the supply rail of the inverter in the current cell.
The wordlines $WL1$ and $WL2$  are shared by each column, and the sourceline $SL$ is shared by a row facilitate the operation of the 2FeFET MIBO. 

The write scheme and the inhibition scheme of the 2FeFET-2T SEE-MCAM design are similar to the 2FeFET-1T SEE-MCAM. During the search operation, the ML state of the current cell is formulated as below:
\begin{equation}
    ML_{i} = ML_{i-1}\times \overbar{D}
\end{equation}
\autoref{fig:prefree}(b) shows the transient waveforms of the proposed 2FeFET-2T SEE-MCAM with 3bit density, validating the MCAM functionality.  

The 2FeFET-2T SEE-MCAM design eliminates the need for precharge in most cases since the ML state $ML_{i}$ is determined by both  the output of 2FeFET MIBO structure $D$ and the previous cell's ML voltage $ML_{i-1}$. 
In consecutive searches,  the ML state of the previous cell $ML_{i-1}$ only changes when 
a mismatch case  in the last search discharges $ML_{i-1}$, and then a match in the current search operation charges $ML_{i-1}$ again.
If consecutive searches yield the same match/mismatch state of the previous cell, $ML_{i-1}$ remains unchanged. Precharging of the current cell $C_i$ occurs only when two conditions are met: (i) the supply rail of $C_i$, $ML_{i-1}$, transitions from  a mismatch state (Low) to a match state (High) during a search, activating the PMOS of $C_i$ to charge $ML_{i}$, and (ii) the MLs of all previous $i-1$ cells are at a match state, enabling a charging path from the voltage supply to $ML_i$. These strict conditions significantly reduce the chances of charging, resulting in  much lower energy consumption. However, it is important to note that the NAND type ML connection introduces higher latency as the sense amplifier  needs to wait for the ML state transition to propagate through the entire word.  


\section{Evaluation \& Benchmarking}
\label{sec:eval}
\begin{table*}[]
\centering
\caption{Comparisons of   CAM Designs}
\label{tab:comparison}
\resizebox{2\columnwidth}{!}{
\begin{tabular}{|c|ccccccc|}
\bottomrule
\textbf{Designs} & \textbf{Device}& \textbf{Cell Structure}\eat{&\textbf{Matching scheme}}& \textbf{Type}&\textbf{Search energy per bit $(fJ)$} & \textbf{Latency $(ps)$}& \textbf{Area per bit ($\mu m^2$)}& \textbf{NVM/MOS node $(nm)$}   \\\toprule\bottomrule
\textbf{16T CMOS}~\cite{pagiamtzis2006content}&CMOS&16T&BCAM&0.59 ($\times9.8$)&582.4 ($\times1.6$)&1.12 ($\times9.3$)&-/45\\
\textbf{DAC'22}~\cite{cai2022energy}&FeFET&2T-1FeFET&BCAM&0.116 ($\times1.9$)&401.4 ($\times1.1$)&0.36 ($\times3$)&45/45
\\
\textbf{Nat Ele'19}~\cite{ni2019ferroelectric}&FeFET&2FeFET&TCAM&0.40 ($\times6.7$)&360 ($\times1$)&0.15 ($\times1.2$)&45/-
\\
\textbf{DATE'21 (P$^\#$)}~\cite{qian2021energy}&FeFET&2FeFET-1T\eat{&Exact}&TCAM&0.195 ($\times3.3$)&252.8 ($\times0.7$)&0.36 ($\times3$)& 45/45
\\
\textbf{DATE'21 (PF$^\#$)}~\cite{qian2021energy}&FeFET&2FeFET-2T\eat{&Exact}&TCAM&0.073 ($\times1.2$)&1430 ($\times3.8$)&0.44 ($\times3.6$)& 45/45
\\
\textbf{JSSC'13}~\cite{li20131}&PCM&2T-2R\eat{&Exact}&TCAM&0.55 ($\times9.2$)&350.6 ($\times0.9$)&0.41 ($\times3.4$)&90/90
\\
\textbf{NC'20}~\cite{li2020analog}&ReRAM&6T-2R\eat{&Threshold}&ACAM&0.52 ($\times8.7$)&110 ($\times0.3$)&0.51$^\P$ ($\times4.2$)&50/180
\\
\textbf{TED'20}~\cite{yin2020fecam}&FeFET&2FeFET\eat{&Exact/Threshold}&MCAM/ACAM&0.182/0.069 ($\times3$/$\times1.2$)&-&0.05($\times0.4$)&45/45
\\
\textbf{IEDM'20}~\cite{li2020scalable}&FeFET&2FeFET-1T$^\ddagger$\eat{&Exact}&MCAM&0.292 ($\times4.9$)&422 ($\times1.1$)&0.03$^\&$ ($\times0.2$)& 28/-
\\
\textbf{This work (P)}&FeFET&2FeFET-1T&MCAM&0.06$^\dagger$ ($\times1$)&371.8 ($\times1$)&0.12 ($\times1$)& 45/40
\\
\textbf{This work (PF)}&FeFET&2FeFET-2T\eat{&Exact}&MCAM& 0.039$^\dagger$&2040 &0.146& 45/40
\\
\toprule
\end{tabular}
}
\flushleft
$\dagger$: Results are evaluated under 32 cells per word. 
$\ddagger$: Two ML branches for one stored vector. \eat{$*$: CAM with novel distance functions. $\star$: Single-cell metrics are not reported.} $\#$: Precharge and precharge-free. 
$\P$: Reported based on 16nm design rules.
$\&$: Smaller cell size due to 28nm technology node.

\end{table*}

In this section, we evaluate and compare the search energy per bit and delay of the proposed SEE-MCAM arrays with existing BCAM/TCAM and MCAM designs to validate the benefits of exploiting MLC FeFET devices and energy efficient design schemes. We then benchmark the proposed design in the context of a quantized HDC model at application level.

\subsection{SEE-MCAM Evaluations}
\label{sec:eval}
Our proposed SEE-MCAM designs utilize
the 45nm Preisach FeFET model \cite{ni2018circuit}. Different threshold voltage states corresponding to multi-bit values are written to FeFETs via different write pulses \cite{yin2020fecam,li2020scalable}. 
The 40nm UMC processing development kit (PDK) is adopted for all CMOS transistors. 
All the  circuits have been simulated with Cadence. Wiring parasitics associated with MLs are extracted from DESTINY \cite{poremba2015destiny}. 

\begin{figure}
    \centering
    \includegraphics[width=\linewidth]{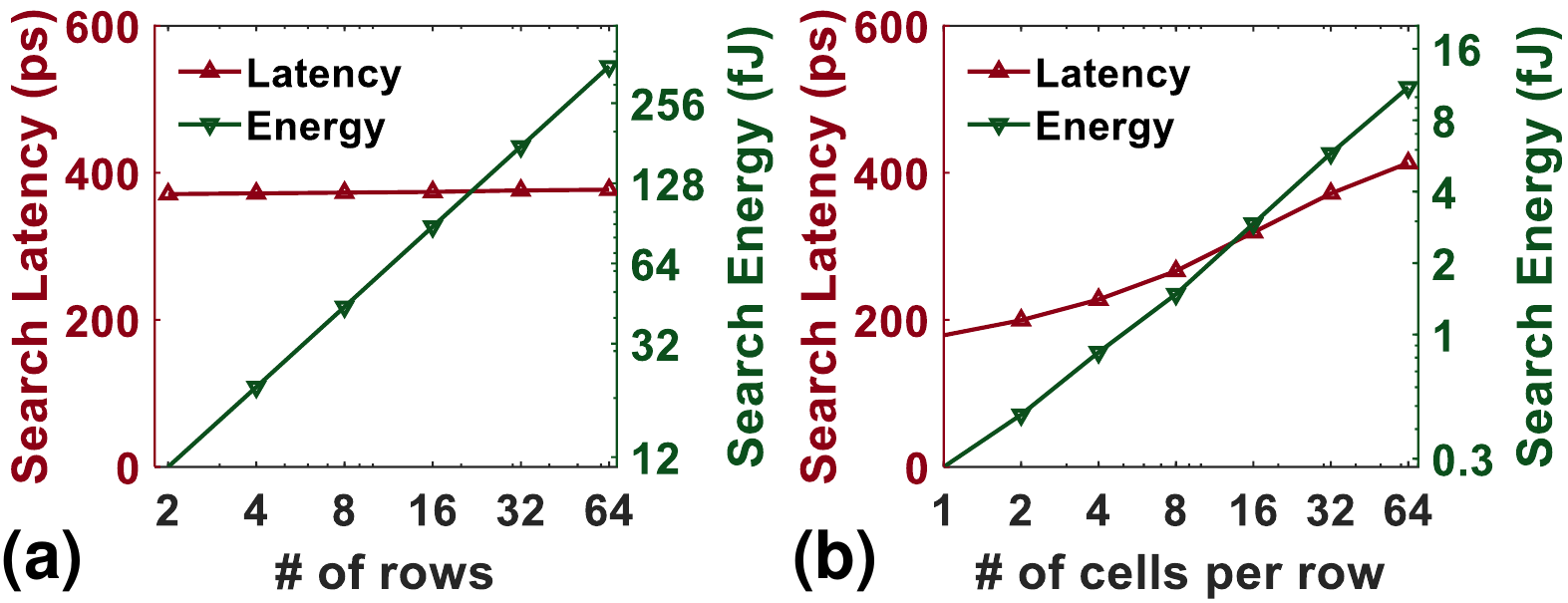}
    \caption{Search latency and  energy of 2FeFET-1T \design (a) with  varying number of rows; (b) with  varying number of cells per row.}
    \label{fig:scalability2}
\end{figure}
\begin{figure}
\centering
    \includegraphics[width=\linewidth]{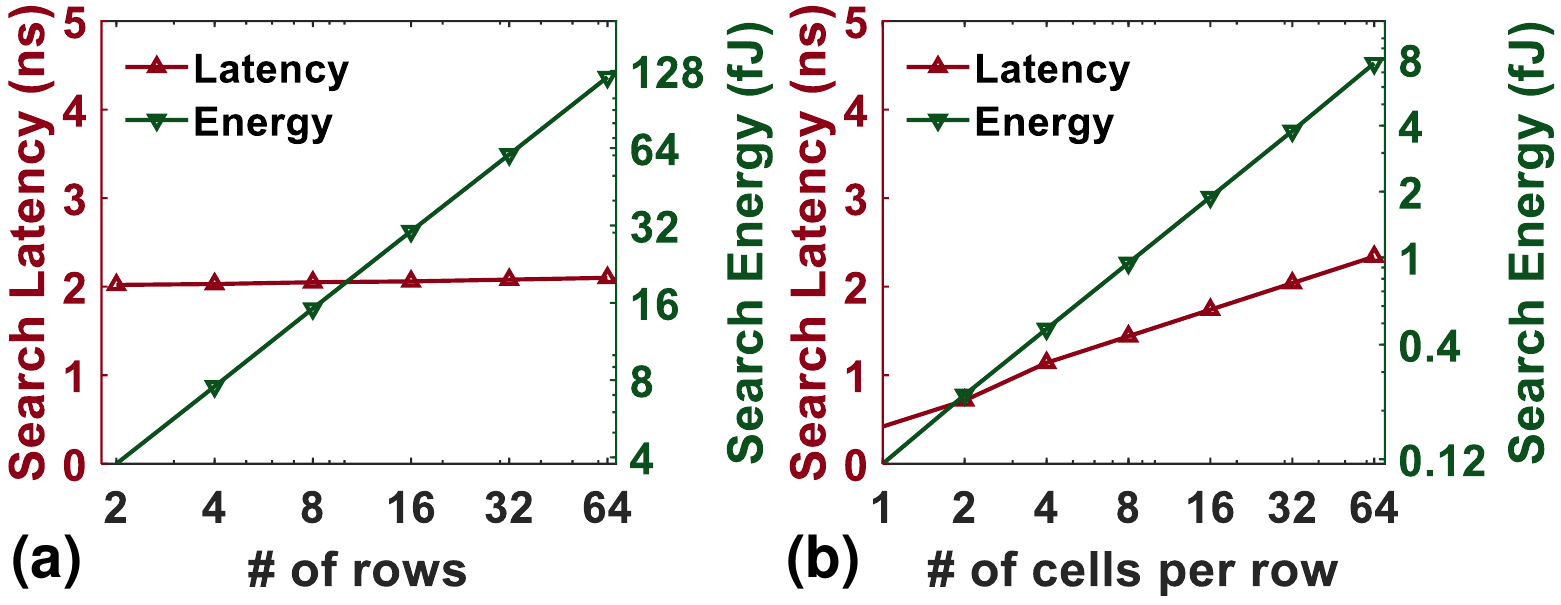}
    \caption{Search latency and  energy of 2FeFET-2T \design (a) with  varying number of rows; (b) with  varying number of cells per row.}
    \label{fig:scalability_free}
\end{figure}

\autoref{fig:scalability2} and \autoref{fig:scalability_free} show the search energy and latency of the proposed SEE-MCAM arrays with varying number of rows and number of cells per row. The latency is reported under the worst case, i.e. one mismatch. It can be seen from \autoref{fig:scalability2}(a) and \autoref{fig:scalability_free}(a) that the respective search energy of both 2FeFET-1T and 2FeFET-1T SEE-MCAM arrays increase linearly as the number of rows increases.  
As the rows of both arrays are independent of each other, the search latency only slightly changes with the number of rows.
On the other hand, it can be seen from \autoref{fig:scalability2}(b) and \autoref{fig:scalability_free}(b) that, as the number of cells increases, the associated ML capacitance increases, slowing down the precharge/discharge and signal propagation speed, and resulting in an increase of search latency. The increasing number of cells also leads to larger precharge energy associated with ML capacitance, as well as larger load energy associated with SLs.

\autoref{tab:comparison} summarizes and compares different NVM based CAM designs with our proposed SEE-MCAM designs  in terms of search energy per bit, latency and area overhead per bit. The cell sizes are estimated based on a 2X2 SEE-MCAM array layout. The area per bit of our proposed 2FeFET-1T SEE-MCAM is 8\% of the conventional 16T CMOS CAM. Moreover, the proposed 2FeFET-1T SEE-MCAM achieves 9.8$\times$ more energy efficiency and 1.6$\times$ less search latency than CMOS CAM, respectively.
Regarding the NVM-based counterparts, on one hand, multi-bit CAM functionality leads to much higher energy efficiency compared to the BCAM and TCAM designs.
For example, our proposed 2FeFET-1T SEE-MCAM design is 6.7$\times$ more energy efficient than the typical 2FeFET TCAM design \cite{ni2019ferroelectric}, respectively. 
On the other hand, the ML capacitance reduction and precharge-free design schemes applied to the SEE-MCAM designs also brings significant energy saving and speedup when
compared to other MCAM designs. Our approach can achieve 8.7$\times$ and 4.9$\times$ more energy efficiency than ReRAM based \cite{li2020analog} and FeFET-based  \cite{li2020scalable} MCAM designs, respectively.
Though ReRAM-based MCAM consumes less search latency, this is due to the high sensing current of 6T-2R cell structure. 
Overall, these evaluation results validate the efficiency of our SEE-MCAM approaches for associative search applications.

\begin{figure}
    \centering
    \includegraphics[width=\linewidth]{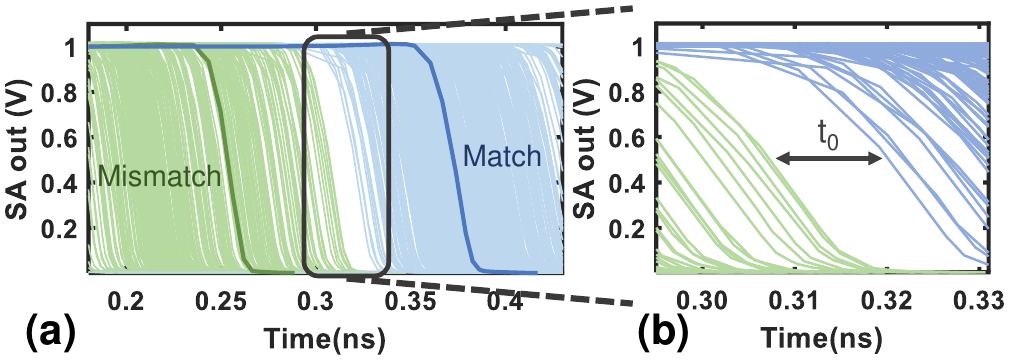}
    \caption{Transient waveforms of the proposed  SEE-MCAM array with device variability under the worst case.}
    \label{fig:MC}
\end{figure}

We also validate the robustness of our proposed \design design and the ML scheme. 
The variations of all the CMOS transistors, including the sense amplifier, are modeled using the PDK with {\tt{TT}} process corner at a temperature of \ang{27}C. 
The   variations of FeFET devices are obtained from experimentally measured devices, with a standard deviation $\sigma=54mV$ for the low/high $V_{TH}$ state \cite{soliman2020ultra}.
Additionally,  smaller FeFET variation can be achieved through the use of a write-and-read verify scheme \cite{hu2021memory}.
\autoref{fig:MC} depicts the transient waveforms of the SA output including process variations during the search operations in the proposed SEE-MCAM array.
The transient results of 100 Monte Carlo simulations shown in \autoref{fig:MC} demonstrate sufficient sense margin of our proposed design even in the worst search case, confirming  the robustness of our proposed design against device variations. 

\begin{figure}
    \centering
    \includegraphics[width=\linewidth]{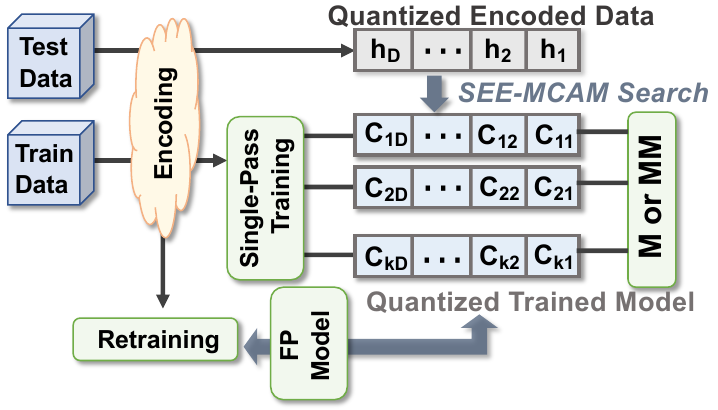}
    \caption{Overview of the quantized HDC framework leveraging the proposed \design design.}
    \label{fig:HDC_overview}
\end{figure}

\subsection{Benchmarking: Quantized Hyperdimensional Computing}
\label{sec:HDC}


We further benchmark our proposed \design as an associative memory in the context of a novel hyperdimensional computing (HDC) framework. 
Inspired by the fact that the brain computes based on patterns that are not readily related to numbers, HDC is built upon a set of transparent operations and is extremely robust to hardware noise due to the \textit{holographic} and \textit{redundant} nature. It has been proposed as an alternative computing paradigm for resource-constrained edge scenarios.

Previous IMC designs \cite{liu2022cosime} are benchmarked with a full precision HDC model \cite{hernandez2021onlinehd}. However, due to the limited precision of IMC designs, assuming full precision for benchmarking may lead to less accurate results.
In this work, we employ our approach in a calibrated quantized HDC model to conduct a more practical benchmarking. 
The framework is implemented in Python  with Pytorch packages and supports both full precision and quantized HDC inference. \autoref{fig:HDC_overview} shows the overall quantized HDC framework. 
Non-linear quantization is performed on the encoded query and the class hypervectors that are stored in the \ design-based associative memory.
The multi-bit exact match scheme of \design is adopted. 
Here, we first discuss the basics of the quantized HDC.

\textbf{Encoding:} HDC encoding refers to mapping the feature from low dimensional space 
$\mathcal{F}\subset\mathbb{R}^d$ into high dimensional space 
$\mathcal{H}\subset\mathbb{R}^D$ where dimensionality $D>>d$. 
For instance, a vector $\vec{F} = [f_1, ..., f_n]$ with n features is multiplied with an $n\times D$ matrix $\vec{B}$, where every element in $\vec{B}$ is sampled from i.i.d Gaussian distribution with $\mu=0$ and $\sigma=1$.

\textbf{Training and Retraining:}
The lightweight training of HDC often refers to single-pass training, which is amenable to edge devices. Hypervector $\vec{H_l}$ associated to the label $l$ is generated after the encoding phase. 
For single-pass training in \autoref{fig:HDC_overview}, all the $\vec{H_l}$ are aggregated ($k$ $\vec{H_l}$s is presented here): $\vec{C_l} = \sum_k \vec{H_l}$.
For iterative training, HDC trains the data as follows:
\begin{equation}
\label{eq:train}
  \begin{aligned}
    \vec{C_l} & \leftarrow \vec{C_l} + \eta(1-\delta)\vec{Q}\\
    \vec{C_{l^{\prime}}} & \leftarrow \vec{C_{l^{\prime}}} - \eta(1-\delta)\vec{Q}.
  \end{aligned}
\end{equation}
Where $\eta$ is the learning rate set to 0.03 in this work, $l^{\prime}$ is the mispredicted label, and $l$ is the correct label.

\textbf{Inference:} After the hypervectors are generated via the encoding phase and trained based on \autoref{eq:train}, they are then stored in the SEE-MCAM array for inference.
When a new query $\vec{Q}$ comes in for classification, it will first be encoded in the encoding phase, and then searched in the associative memory by applying multi-bit voltages corresponding to the element values to the SEE-MCAM array.

In this work, we adopted the quantized HDC model where there exists a full-precision model for  training and a quantized model stored in the SEE-MCAM for inference. 
We quantize each element of a hypervector to the desired bit precision based on its Z-score ($Z = \frac{x-\mu}{\sigma}$) over the Gaussian distribution. Take the proposed 3-bit \design as an example, element values that drop beneath 12.5\% of the cumulative distribution function (CDF) will be assigned to `000'. 

\subsubsection{Classification Accuracy}
\begin{figure}
    \centering
    \includegraphics[width=\linewidth]{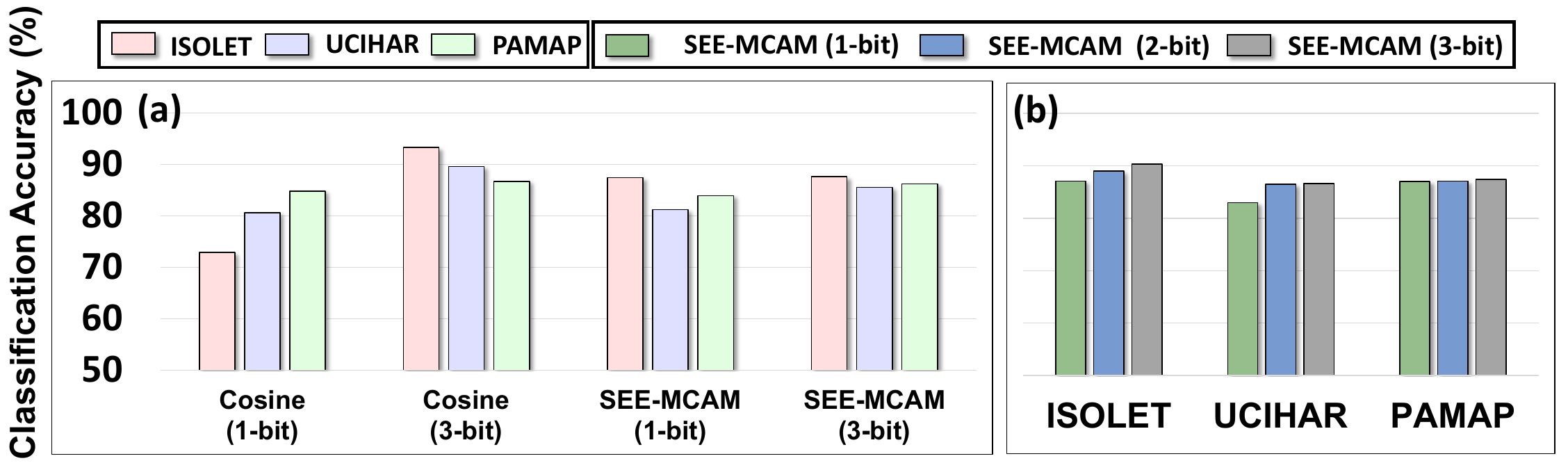}
    \caption{Quantized HDC benchmarking accuracy with (a)    binary cosine similarity, 3-bit cosine similarity, binary \design, and 3-bit \design under $D = 1024$, respectively; 
    (b) \design implementation supporting varying dimensionality ($D=1024$, $D=2048$, and $D=4096$, respectively).}
    \label{fig:bench1}
\end{figure}
\begin{table}[]
\centering
\caption{Datasets ($n$: feature size, $K$: number of classes)}
\label{tab:benchmark}
\resizebox{1\columnwidth}{!}{
\begin{tabular}{|c|ccccc|}
\toprule
\textbf{Dataset}& $n$ & $K$ & \shortstack{\textbf{Train}\\ \textbf{Size}} & \shortstack{\textbf{Test}\\ \textbf{Size}} & \textbf{Description}             \\ \midrule
\textbf{ISOLET} & 617               & 26         & 6,238               & 1,559              & Voice Recognition~\cite{Isolet}              \\
\textbf{UCIHAR} & 561               & 12          & 6,213                & 1,554               & Physical Activity Monitoring~\cite{anguita2012human} \\ 
\textbf{PAMAP} &  75             &  5        &  611,142  & 101,582           & Human Activity Recognition~\cite{reiss2012introducing}           \\
\bottomrule
\end{tabular}
}
\end{table}
\autoref{fig:bench1}(a) illustrates the quantized HDC accuracy on three different datasets, whose descriptions are summarized in \autoref{tab:benchmark}. 
As HDC typically exploits cosine distance between hypervectors as the optimal similarity function during the inference, we hereby compare the our SEE-MCAM based quantized HDC implementation with the quantized  HDC framework based on cosine distance. Both implementations quantize the elements of hypervectors after training to 3 bits, respectively, using the aforementioned non-linear quantization scheme. 
Inference accuracy results shown in \autoref{fig:bench1} indicate that the proposed 3-bit \design based implementation has on average 3.43\% accuracy degradation compared to the 3-bit cosine similarity-based implementation in GPU.  
To make a fair comparison, 
We also implement COSIME, a binary cosine similarity-based associative memory \cite{liu2022cosime}, in our quantized HDC framework, and compare it with our proposed SEE-MCAM based implementation. 
The result shown in \autoref{fig:bench1}(a) indicates that the binary SEE-MCAM based implementation achieves on average 2.26\% accuracy improvement over COSIME-based implementation.

Moreover, as the proposed \design increases the data density  compared to BCAM/TCAM, the SEE-MCAM based HDC implementation can actually implement higher dimensionality without extra hardware cost. 
With the same amount of CAM cells employed (e.g. 1024) in the HDC framework, \design can represent and store much more elements per hypervector (e.g. $D=2048$ for 2-bit \design and $D=4096$ for 3-bit \design).
For HDC, increasing the dimensionality of the hypervector leads to higher algorithmic accuracy.
As a result, the 3-bit SEE-MCAM achieves on average 2.41\% accuracy  improvement over the binary \design as shown in \autoref{fig:bench1}(b), showcasing the superiority of the proposed SEE-MCAM over existing BCAM/TCAM. 

\subsubsection{Hardware Acceleration}
\begin{figure}
    \centering
    \includegraphics[width=\linewidth]{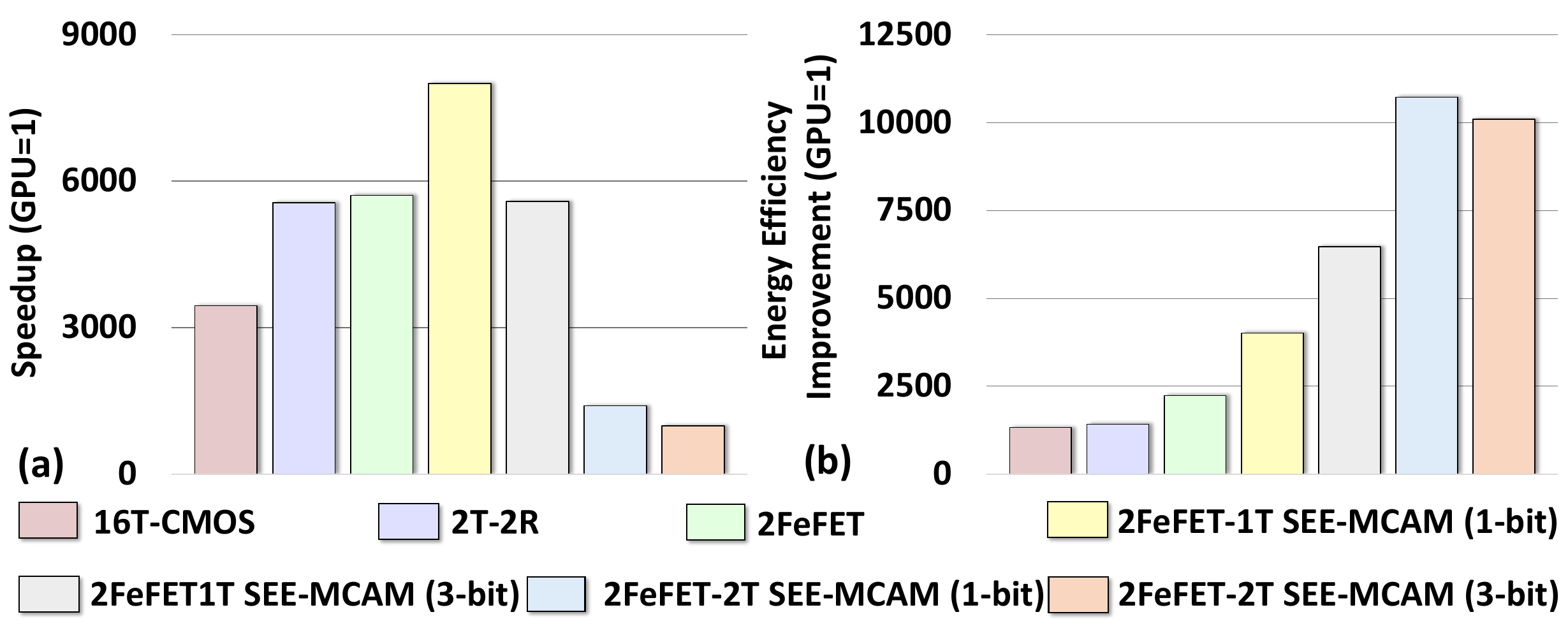}
    \caption{(a) Computational speedup and (b) energy efficiency improvement of \design  compared to a GPU implementation.}
    \label{fig:bench2}
\end{figure}

We further investigate the speedup and energy efficiency improvement of the  SEE-MCAM based quantized HDC implementation over the Nvidia Pascal microarchitecture GTX 1080ti GPU. 
Nvidia System Management Interface is used for accurate power consumption measurement, and Pytorch profiler is used for algorithmic delay breakdown. The profiler delay for exact matching is extracted from Pytorch Aten's API.

We compare the proposed \design with various CAM designs, including 16T CMOS \cite{pagiamtzis2006content}, 2T-2R \cite{li20131}, 2FeFET \cite{ni2019ferroelectric}, in the context of the quantized HDC framework. All results are respective to the same tasks running on the GPU.
Both binary and 2-bit \design designs have been incorporated into the HDC framework and evaluated.
It can be seen from \autoref{fig:bench2} that our proposed approach offers up to {3} orders of magnitude  speedup and energy efficiency improvement than GPU. 
As shown in \autoref{fig:bench2}(b),  the proposed \design implementations significantly improve the energy efficiency than other CAM-based implementations. 
Both \autoref{fig:bench2} (a) and (b) illustrate the benefit trend of BCAM, TCAM and MCAM designs over a GPU implementation.
\section{Conclusion}

In this work, we propose \design,  scalable energy-efficient MCAM designs that exploit FeFETs as proxy to improve the CAM density and energy efficiency over existing BCAM/TCAM and MCAM designs. We leverage the MLC property of FeFETs to construct a 2FeFET MIBO structure, which is the key part of SEE-MCAM. We then propose NOR type 2FeFET-1T and NAND type 2FeFET-2T SEE-MCAM designs to implement multi-bit CAM functionality and achieve significant energy efficiency at the same time.
The functionality and robustness of the proposed approaches have been validated. Evaluation results at array level and quantized HDC application benchmarking suggest that our proposed SEE-MCAM designs achieves better data density, energy efficiency and performance when compared with other state-of-the-art CAM designs.

\section*{Acknowledgements}
This work was supported in part by  Zhejiang Provincial Natural Science Foundation (LD21F040003, LQ21F040006), NSFC (62104213, 92164203), National Key Research and Development Program of China (2022YFB4400300). Liu and Wan were supported by COCOSYS, one of seven centers in JUMP2.0, a SRC program sponsored by DARPA. This work was supported in part by National Science Foundation \#2127780 and \#2312517, Semiconductor Research Corporation (SRC), and generous gifts from Xilinx and Cisco.

{

\bibliographystyle{IEEEtran}
\bibliography{bib}
}
\end{document}